\documentclass[12pt]{article}
\usepackage{graphicx}

\def\leaderfill{\leaders\hbox to 1em{\hss.\hss}\hfill}  

\begin{document}

\noindent {\footnotesize\it  ISSN 1063-7737,
 Astronomy Letters, 2008, Vol. 34, No. 8, pp. 515--528.
 \copyright Pleiades Publishing, Inc., 2008.
 \noindent Original Russian Text
 \copyright V.V. Bobylev, A.T. Bajkova, A.S. Stepanishchev, 2008,
 published in Pis'ma v Astronomicheski$\check{\imath}$ Zhurnal,
 2008, Vol. 34, No. 8, pp. 570--584.}

\noindent
\begin{tabular}{llllllllllllllllllllllllllllllllllllllllllllll}
& & & & & & & & & & & & & & & & & & & & & & & & & & & & & & & & & & & & &  \\
\hline \hline
\end{tabular}

\vskip 1.5cm

 \centerline {\large\bf  Galactic Rotation Curve and the Effect of Density Waves}
 \centerline  {\large\bf from Data on Young Objects}
 \bigskip
 \centerline {V.V. Bobylev, A.T. Bajkova, and A.S. Stepanishchev,}
 \bigskip
 \centerline {\small\it
Central (Pulkovo) Astronomical Observatory of RAS, St-Petersburg}
 \bigskip

{\bf Abstract--}Based on currently available data on the
three-dimensional field of space velocities of young ($\leq$50
Myr) open star clusters and the radial velocities of HI clouds and
star-forming (HII) regions, we have found the Galactic rotation
curve in the range of Galactocentric distances 3 kpc $<R<12$ kpc
using the first six terms of the Taylor expansion of the angular
velocity of Galactic rotation in Bottlinger's equations. The
Taylor terms found at the Galactocentric distance of the Sun
$R_0=7.5$~kpc are:
 $\omega_0=  -27.7\pm0.6$   km s$^{-1}$ kpc$^{-1}$,
 $\omega_0^1=   4.13\pm0.07$  km s$^{-1}$ kpc$^{-2}$,
 $\omega_0^2= -0.912\pm0.065$ km s$^{-1}$ kpc$^{-3}$,
 $\omega_0^3=  0.277\pm0.036$ km s$^{-1}$ kpc$^{-4}$,
 $\omega_0^4= -0.265\pm0.034$ km s$^{-1}$ kpc$^{-5}$,
 $\omega_0^5=  0.104\pm0.020$ km s$^{-1}$ kpc$^{-6}$.
 In this case, the Oort
constants are
 $A= 15.5\pm0.3$ km s$^{-1}$ kpc$^{-1}$ and
 $B=-12.2\pm0.7$ km s$^{-1}$ kpc$^{-1}$.
We have established that the centroid of the sample moves relative
to the local standard of rest along the Galactic $Y$ axis with a
velocity of $-6.2\pm0.8$ km s$^{-1}$. A Fourier spectral analysis
of the velocity residuals from the derived rotation curve
attributable to density waves reveals three dominant peaks with
wavelengths of 2.5, 1.4, and 0.9 kpc and amplitudes of 4.7, 2.6,
and 3.6 km s$^{-1}$, respectively. These have allowed us to
estimate the distances between the density wave peaks, 1.9, 2.4,
and 3.2 kpc as R increases, in agreement with the description of
the density weave as a logarithmic spiral. The amplitude of the
density wave perturbations is largest in the inner part of the
Galaxy, $\approx9$ km s$^{-1}$, and decreases to $\approx1$ km
s$^{-1}$ in its outer part. A spectral analysis of the radial
velocities of young open star clusters has confirmed the presence
of periodic perturbations with an amplitude of $5.9\pm1.1$ km
s$^{-1}$ and a wavelength $\lambda=1.7\pm0.5$ kpc. It shows that
the phase of the Sun in the density wave is close to $-\pi/2$ and
the Sun is located in the interarm space near the outer edge of
the Carina-Sagittarius arm.

\section*{INTRODUCTION}

Constructing the Galactic rotation curve is of great importance in
solving a number of problems, such as estimating the mass of the
Galaxy, determining the distribution of matter, estimating the
hidden mass, studying the dynamics of the Galaxy and its
subsystems, etc. The shape of the rotation curve in the outer
Galaxy is of particular interest. Xue et al. (2008) showed that
the circular rotation velocity is close to 200 km s$^{-1}$ far
from the Galactic center, being 175 km s$^{-1}$ at a distance of
60 kpc.

The following data on various objects in the Galactic disk are
used to determine the Galactic rotation parameters: HI and HII
radial velocities (Burton 1971; Clemens 1985; Fich et al. 1989;
Brand and Blitz 1993; Nikiforov 1999; Russeil 2003; Avedisova
2005), data on open star clusters (OSCs) and associations,
Cepheids and other young stars (Mishurov and Zenina 1999;
Rastorguev et al. 1999; Dambis et al. 2001; Zabolotskikh et al.
2002; Dias and L\'epine 2005).

The most comprehensive information about the pattern of Galactic
rotation in the inner region can be obtained using the radial
velocities of hydrogen clouds extracted from 21-cm or CO radio
observations.

Once the theory of spiral density waves has been developed (Lin
and Shu 1964; Lin et al. 1969), it turned out that the small
deviations of the circular velocities of hydrogen clouds from the
smooth rotation curve constructed from radial velocities could be
successfully interpreted in terms of the density wave theory
(Burton 1971; Burton and Gordon 1978; Clemens 1985). However, the
hydrogen radial velocity data are insufficient to trace the
kinematic effect of the spiral pattern in the outer Galaxy.
(Indeed, since the tangential point method is inapplicable, the
hydrogen data have large errors.) Therefore, other data, for
example, those on OSCs, should be invoked additionally.

When only the radial velocities are analyzed, the following two
quantities serve as model ones for constructing the Galactic
rotation curve: the Galactocentric distance of the Sun $R_0$ and
the linear circular velocity of the Sun $V_0$ (Brand and Blitz
1993; Russeil 2003; Avedisova 2005). Including OSCs for which
reliable estimates of their proper motions are available allows us
either to reduce the number of model quantities to one ($R_0$),
because in this case the angular velocity of solar rotation
$\omega_0$ can be determined from observations (Rastorguev et al.
1999; Zabolotskikh et al. 2002), or to determine both $R_0$ and
$V_0$ directly from observational data (Bobylev et al. 2007).

The goal of this paper is to solve two problems: (1) constructing
a smooth rotation curve from young objects in the Galactic disk
and (2) analyzing the high-frequency periodic circular velocity
residuals of the objects used from the derived rotation curve
attributable to the effect of spiral density waves.

To solve the first problem, the $R$ range is usually broken down
into several segments: the central ($R<0.8-1.5R_0$), inner
($\approx1.5R_0<R<R_0$), and outer ($R>R_0$) regions of the
Galaxy, as was done by Burton (1971) and Clemens (1985). This
approach is used mainly because there is a segment of solid-body
rotation in the inner Galaxy that passes into a fairly flat curve
further out. Therefore, the entire $R$ range is difficult to
describe by a smooth curve. In addition, there are significant
radial motions of gas at $R<3$ kpc (Bania 1977; Liszt and Burton
1980). Several authors suggested their smoothed rotation curves
for each of these regions (Burton and Gordon 1978; Clemens 1985).
However, it is not always convenient to use a rotation curve
defined by different expressions in different $R$ segments. It is
much more convenient to have one analytical expression for the
rotation curve in the entire range.

Here, we restrict our analysis to the range of distances
$R>2.5$~kpc. Zabolotskikh et al. (2002) found a smoothed Galactic
rotation curve in a wide range of distances $R$ without dividing
it into segments, but this rotation curve was given in tabular
form, which complicates its use. Our objective is to determine the
Galactic rotation curve in a wide range of distances, $R>2.5$~kpc,
in analytical form.

We use a Fourier spectral analysis to analyze the velocity
residuals of the objects from the derived rotation curve produced
by a spiral density wave.

Apart from the radial velocities of hydrogen clouds, we invoke
data on star-forming regions (Russeil 2003) and new data on young
OSCs (Kharchenko et al. 2007). To determine the projections of the
OSC circular velocities, we use their total space velocities.

\section*{DATA}
\subsection*{The Working Data Set}

We took the radial velocities of HI clouds obtained by the
tangential point method from Fich et al. (1989), where the 21-cm
hydrogen radio observations described in Burton and Gordon (1978)
are presented in tabular form. We also used the CO radio
observations by the tangential point method from Clemens (1985).

We took the data on star-forming (H II) regions mainly from the
catalog by Russeil (2003), which contains the radial velocities
and photometric distances for 204 regions. In this paper, we use
only 89 HII regions selected according to the following criteria:

(i) the relative random errors in the distances to the objects do
not exceed $20\%$;

(ii) the regions are located outside the zone with the Galactic
center--anticenter direction $|l|<12^\circ$, where the chance of
estimating the radial velocities is reduced sharply (Russeil
2003);

(iii) the regions have Galactocentric distances of more than 5 kpc
(closer regions are rejected because of the large velocity
dispersion).

For the HII region Sharpless 294, we use a new estimate of its
photometric distance, $r=4.8\pm0.2$ kpc, from Samal et al. (2007).
For the star-forming region W3OH, we use a new estimate of its
distance, $r=1.95\pm0.04$ kpc, from Xu et al. (2006) obtained from
the trigonometric parallax ($0.512\pm0.010$ mas) measured using
the VLBA. For the HII region Sharpless 269, we use the results of
high-precision radio interferometry (Honma et al. 2007): the
trigonometric parallax $0.189\pm0.008$ mas ($r=5.28\pm0.23$ kpc),
the radial velocity $V_{LSR}=19.6$ km s$^{-1}$, and the proper
motion components
 $(\mu_\alpha \cos\delta, \mu_\delta)_{J2000} =(-0.422\pm0.010,-0.121\pm0.042)$ mas
determined relative to an extragalactic source. Thus, these data
allow the space velocities of the region Sharpless 269 to be
determined.

For OSCs, we use the coordinates, proper motions, radial
velocities, and age estimates from the compilation by Bobylev et
al. (2007), which is based on the catalog by Piskunov et al.
(2006). We took new radial velocities for a number of OSCs from
the CRVOCA catalog (Kharchenko et al. 2007). In this paper, we
consider only the OSCs whose ages do not exceed 50 Myr and for
which the space velocities can be determined. We took the data for
two young clusters, Cr 272 and Tr 28, which are missing from the
CRVOCA catalog, from the catalog by Dias et al. (2002).

\subsection*{Heliocentric Radial Velocities}

We will work with the radial velocity estimates for the objects
given in the heliocentric frame of reference. Different authors
give their radial velocity estimates relative to the local
standard of rest using differing parameters --- $(U, V, W)_{LSR}$.

We reduce ~~the hydrogen~~ radial velocities from Burton and
Gordon (1978) and Clemens (1985) to the heliocentric frame of
reference using the velocities $(U, V, W)_{LSR} = (10.3, 15.3,
7.7) $ km s$^{-1}$. We calculated these values based on the
following parameters of the standard solar motion:
 $(\alpha, \delta)_{1900}=(270^\circ, +30^\circ)$ and
 $V= 20$ km s$^{-1}$ (Burton and Gordon 1978).

We reduce the radial velocities of the star-forming regions to the
heliocentric frame of references using the velocities
$(U,V,W)_{LSR} = (10.4,14.8,7.3)$ km s$^{-1}$ from Russeil (2003).

We reduce the radial velocities of the HII region Sharpless 269 to
the heliocentric frame of reference using the velocities
$(U,V,W)_{LSR} = (10.0,15.4,7.8)$ km s$^{-1}$ from Honma et al.
(2007).

All proper motions of the objects are given for the epoch J2000.0,
i.e., they are homogeneous. Therefore, no problem with their
reduction to a single frame of reference arises.

\section*{METHODS}

In this paper, we use a rectangular Galactic coordinate system
with the axes directed away from the observer toward the Galactic
center $(l=0^\circ, b=0^\circ$, the $X$ axis), along the Galactic
rotation ($l=90^\circ, b=0^\circ$, the $Y$ axis), and toward the
North Galactic Pole ($b=90^\circ$, the $Z$ axis).

\subsection*{Determining the Rotation-Curve Parameters}

The method used here is based on Bottlinger's well-known formulas
(Ogorodnikov 1958) with the angular velocity of Galactic rotation
expanded in a series to terms of the sixth order of smallness in
$r/R_0$:
$$
\displaylines{\hfill
 V_r=-u_\odot\cos b\cos l -
 \hfill\llap(1)\cr\hfill
     -v_\odot\cos b\sin l
     -w_\odot\sin b-
 \hfill\cr\hfill
 -R_0\sin l \cos b[ (R-R_0) \omega^1_0/1!+
 \hfill\cr\hfill
 +...+(R-R_0)^5  \omega^5_0/5!],
  \hfill\cr\hfill
 V_l= u_\odot\sin l - v_\odot\cos l -
 \hfill\llap(2)\cr\hfill
  -(R_0\cos l - r\cos b) [(R-R_0)\omega^1_0/1!+
 \hfill\cr\hfill
  +...+(R-R_0)^5 \omega^5_0/5!]+
 \hfill\cr\hfill
  + r \omega_0 \cos b,\hfill\cr\hfill
 V_b=u_\odot\cos l \sin b +
 \hfill\llap(3)\cr\hfill
         +v_\odot\sin l \sin b
         -w_\odot\cos b +
  \hfill\cr\hfill
 +R_0\sin l \sin b[(R-R_0)\omega^1_0/1!+
  \hfill\cr\hfill
 +...+(R-R_0)^5\omega^5_0/5!],
\hfill
 }
$$
where $V_r$ is the radial velocity; $V_l = 4.74r\mu_l \cos b$ and
$V_b = 4.74r\mu_b$ are the proper motion velocity components in
the $l$ and $b$ directions, respectively (the coefficient 4.74 is
the quotient of the number of kilometers in the astronomical unit
by the number of seconds in the tropical year); $r$ is the
heliocentric distance of the object; the proper motion components
$\mu_l \cos b$ and $\mu_b$ are in mas yr$^{-1}$, the radial
velocity $V_r$ is in km s$^{-1}$; $u_\odot,v_\odot,w_\odot$ are
the solar velocity components relative to the centroid under
consideration; $R_0$ is the Galactocentric distance of the Sun,
which we take to be $R_0=7.5$ kpc (Bobylev et al. 2007); $R$ is
the Galactocentric distance of the object; $R$, $R_0$, and $r$ are
in kpc. The quantity $\omega_0$ is the angular velocity of
rotation at the distance $R_0$, the parameters
 $\omega^1_0, ..., \omega^5_0$ are the derivatives of the angular velocity from the
first to the fifth order, respectively. The distance $R$ can be
calculated using the expression
$$
\displaylines{\hfill
 R^2=r^2\cos^2 b-2R_0 r\cos b\cos l+R^2_0.\hfill\llap(4)
 }
$$
Note also that Eqs. (1)--(3) are written in such a way that the
direction of rotation from the $X$ axis to the $Y$ axis is
positive. The system of conditional equations (1)--(3) contains
nine unknowns: $u_\odot$,$v_\odot$,$w_\odot$,
 $\omega_0,\omega^1_0,\omega^2_0$,
 $\omega^3_0,\omega^4_0,\omega^5_0$,
which can be determined by the least-squares method. Equations
(1)--(3) can be easily generalized to determine any necessary
number of terms of the series in the expansion of the angular
velocity of rotation $\omega^n_0$. The system of equations
(1)--(3) is solved with weights of the form
$$
\displaylines{\hfill
 P_{(r, l, b)}=S_0/\sqrt{S_0^2+\sigma^2_{V_{r, l, b} }},\hfill
 }
$$
where $P_r$ is the weight of the equation for the radial velocity,
$P_l$ and $P_b$ are the corresponding weights for the components
$V_l$ and $V_b$, and $S_0$ denotes the dispersion averaged over
all observations, which has the meaning of the ``cosmic''
dispersion that we take to be 8 km s$^{-1}$. The errors
$\sigma_{V_l}$ and $\sigma_{V_b}$ in the velocities $V_l$ and
$V_b$ can be calculated using the formulas
$$
\displaylines{\hfill
 \sigma_{(V_l, V_b)} = 4.74 r
 \sqrt{\mu^2_{l, b}\Biggl({\sigma_r\over r}\Biggr)^2+\sigma^2_{\mu_{l, b}}},\hfill  
 }
$$
where $\sigma_\pi/\pi$ is taken to be 0.2, a typical error in the
photometric distance for OSCs.

\subsection*{Determining the Projections $V_{rot}$ and $V_{rad}$}

When only the radial velocity is available for an object, we
calculate the projection of the circular rotation velocity
$V_{rot}$ from the well-known formula (Burton 1971)
$$
\displaylines{\hfill
  V_{rot}=|R\omega_0|+RV_r/(R_0\sin l\cos b).\hfill\llap(5)
}
$$
For OSCs, we have the rectangular coordinates $X,Y,Z$ and
calculate the components of the observed space velocities $U$ and
$V$ from the projections $V_r,Vl,$ and $V_b$ (Kulikovskii 1985).
Using these components, we find two projections, $V_{rad}$
directed radially away from the Galactic center and $V_{rot}$
orthogonal to it, from the relations
$$
\displaylines{\hfill
  V_{rot}= U\sin \theta+(V_0+V)\cos \theta, \hfill\llap(6)\cr\hfill
  V_{rad}=-U\cos \theta+(V_0+V)\sin \theta, \hfill\llap(7)
}
$$
where $V_0=|R_0\omega_0|$ and the position angle $\theta$ is
defined as $\tan\theta=Y/(R_0-X)$. We assume that the velocities
$U$ and $V$ are free from the solar velocity relative to the
centroid $V_\odot(u_\odot,v_\odot,w_\odot)$ derived above from
Eqs. (1)--(3). The errors in the projections $V_{rad}$ and
$V_{rot}$ can be estimated from the relations
$$
\displaylines{\hfill
  \sigma^2(V_{rot})= \sigma^2_U\sin^2 \theta+\sigma^2_V\cos^2 \theta, \hfill\cr\hfill
  \sigma^2(V_{rad})= \sigma^2_U\cos^2 \theta+\sigma^2_V\sin^2 \theta, \hfill
}
$$
where the errors in the space velocities $U$ and $V$ are denoted
by $\sigma_U$ and $\sigma_V$.

\subsection*{Analysis of the Residual Velocities}

At the next step, we consider the rotation velocities $\Delta V$
that are the residual ones with respect to the Galactic rotation
curve found at the first step.

A spectral analysis consists in applying the direct Fourier
transform to the sequence of residual velocities in the following
way:
$$
\displaylines{\hfill
 \overline{\Delta V}(\lambda_k)=
 \frac{1}{M}\sum_i^M \Delta
 V^i\exp\biggl(-j\frac{2\pi}{\lambda_k} R_i \biggr),\hfill\llap(8)
}
$$
where $\overline{\Delta V}(\lambda_k)$ is the value of harmonic
$k$ of the Fourier transform, $M$ is the number of measurements of
the residual velocities $\Delta V^i$ with coordinates $R_i$,
 $i=1,2,...,M$, and $\lambda_k$ is the wavelength in kpc; the latter is equal
to $D/k$, where $D$ is the period of the original sequence in kpc.

\subsection*{Constraints}

The constraints listed below reduce significantly the error of a
unit weight when the system of equations (1)--(3) is solved by the
least-squares method.

For HII regions: (i) the error in the distance is no more than
20\%; (ii) $R>5$ kpc; (iii) location outside the zone with the
Galactic center.anticenter direction $|l|<12^\circ$.

For OSCs:
 (i) the age is younger than 50 Myr;
 (ii) the error in the modulus of the space velocity is
 $\sqrt{\sigma^2_{R_V}+\sigma^2_{V_l}+\sigma^2_{V_b}}<15$ km s$^{-1}$;
 (iii) $V_{pec}=\sqrt{U^2+V^2+W^2}<100$ km s$^{-1}$. In
particular, we also use condition (i), because the negative
K-effect, i.e., the effect of radial motions (Bobylev et al.
2007), for young OSCs is less pronounced than that for older ones.

For neutral hydrogen (HI): (i) $R>2.4$ kpc; (ii) several radial
velocities from Burton and Gordon (1978) are not used; these data
are marked by the circles in Figs. 3a and 4. At the same time,
when analyzing the residual velocities, we rejected only two
points close to the Sun. Condition (i) is used, because a
transition to the zone of solid-body rotation is observed on the
rotation curve at $R\approx2.4$ kpc (Burton 1971; Clemens 1985).

The system of conditional equations (1)--(3) is solved twice with
an analysis of the residuals using the $3\sigma$ criterion (we
clearly see from Fig. 4 that the points rejected by this criterion
have large residuals).

\section*{RESULTS}

At a fixed Galactocentric distance of the Sun ($R_0=7.5$ kpc), by
solving the system of 777 equations, we found the parameters of
the solar velocity relative to the centroid,
$(u_\odot,v_\odot,w_\odot) = (8.41,11.41,8.01)\pm(0.49,0.50,0.64)$
km s$^{-1}$, and the parameters of the angular velocity of
Galactic rotation,
$$
\displaylines{\hfill
  ~\omega_0= -27.67\pm 0.61~\hbox{km s$^{-1}$ kpc$^{-1}$},\hfill\llap(9)\cr\hfill
 \omega^1_0= +4.132\pm0.072~\hbox{km s$^{-1}$ kpc$^{-2}$},~\hfill\cr\hfill
 \omega^2_0= -0.912\pm0.065~\hbox{km s$^{-1}$ kpc$^{-3}$},~\hfill\cr\hfill
 \omega^3_0= +0.277\pm0.036~\hbox{km s$^{-1}$ kpc$^{-4}$},~\hfill\cr\hfill
 \omega^4_0= -0.265\pm0.034~\hbox{km s$^{-1}$ kpc$^{-5}$},~\hfill\cr\hfill
 \omega^5_0= +0.104\pm0.020~\hbox{km s$^{-1}$ kpc$^{-6}$}.~\hfill
}
$$
As a result, taking into account all constraints, we used the data
for 140 young OSCs (including S269) with a mean age of 18 Myr, 89
radial velocities of the HII regions (Russeil 2003), 135 radial
velocities for the tangential points (CO) from Clemens (1985), and
133 radial velocities for the tangential points (H I) from Burton
and Gordon (1978). The error of a unit weight is $\sigma_0=8.0$ km
s$^{-1}$.

As a result, the Oort constants
 $A=0.5R_0\omega^1_0$,
 $B=\omega_0+0.5R_0\omega^1_0$,
are
 $A= 15.50\pm0.27$ km s$^{-1}$ kpc$^{-1}$ and
 $B=-12.17\pm0.67$ km s$^{-1}$ kpc$^{-1}$.

Based on parameters (9), we estimated the circular rotation
velocity of the solar neighborhood to be
$V_0=|R_0\omega_0|=208\pm5$ km s$^{-1}$ and the period of its
revolution around the Galactic center to be
 $T=2\pi/(\gamma~\omega_0)=221$~Myr, where the coefficient
 $\gamma=1.023\times10^{-9}$ [km s$^{-1}$ kpc$^{-1}$]/[rad yr$^{-1}$] (Murray 1986).

Figure 1 shows the coordinates of the objects being analyzed in
projection onto the Galactic $XY$ plane.

Figure 2 shows the Galactic rotation curve constructed from
parameters (9); the OSC circular velocity projections were
calculated only from radial velocities using Eq. (5). As can be
seen from Eq. (5), the division by $\sin l$ is made in the second
term. Hence, the projections of the objects with $l$ close to zero
have large deviations and large errors. In Fig. 2, the open
squares mark 17 OSCs with $|l|<12^\circ$. Therefore, Eq. (5) is
not used to analyze the OSC velocities.

Figures 3a and 3b show the OSC circular velocity projections
calculated using Eq. (6). The thick solid line in these figures
indicates the derived rotation curve. Figures 3a and 3b differ by
the age of the OSCs used, whose velocities are denoted by the
circles. Figure 3a pertains to young OSCs (younger than 50 Myr),
while Fig. 3b pertains to old OSCs (older than 50 Myr). We see
from the figure that the velocities of both young and old OSCs
fall nicely on the derived Galactic rotation curve; in contrast to
the old OSCs, the young OSCs exhibit a distinct high-frequency
component related to the effect of density waves. The dotted line
in Fig. 3a indicates the Galactic rotation curve from Zabolotskikh
et al. (2002), which was found using seven terms in the expansion
of the angular velocity. From the figure, we see good agreement
between the two rotation curves for $R>3.5$~kpc.

We solved Eqs. (1)--(3) using various numbers of terms in the
expansion of $\omega_0$. A direct construction of the rotation
curves shows that adding each succeeding term extends
significantly the boundaries of the confidence interval. However,
it can be said with confidence that three expansion terms are
clearly not enough to construct the Galactic rotation curve in the
range of distances 3 kpc$<R<12$ kpc (Bobylev et al. 2007), while
seven terms are too many. For example, when the seventh term is
added to Eqs. (1)--(3), terms that do not differ significantly
from zero appear:
 $\omega^3_0= +0.107\pm0.086$~km s$^{-1}$ kpc$^{-4}$ and
 $\omega^6_0= -0.076\pm0.035$~km s$^{-1}$ kpc$^{-7}$, with the shape of the curve
remaining virtually unchanged. In this sense, we found an optimal
solution.

We can see from our comparison of the OSC circular velocities in
Figs. 2 and 3a that the projections calculated using Eq. (6) have
a smaller residual velocity dispersion with respect to the derived
rotation curve.

The dotted line in Fig. 3b denotes the rotation curve as derived
by Dias and L\'epine (2005) for $R_0=7.5$~kpc; we added a constant
term of 18 km s$^{-1}$ to $V_0=190$ km s$^{-1}$ to reconcile it
with our data at $R=R_0(V_0=208$ km s$^{-1}$).

Figure 4 shows the angular velocity of Galactic rotation
determined using parameters (9). The circles mark the HI data
points rejected by the $3\sigma$ criterion when seeking for the
rotation curve. Note, however, that these points were not rejected
at the second step during our spectral analysis of the
high-frequency deviations. Based on Fig. 4, we choose the range of
distances 3 kpc $<R<12$ kpc as the most significant one.

The $R$ distribution of the residual velocities obtained by
subtracting the derived Galactic rotation curve from the original
distribution of circular velocities is shown in Fig. 5. This
distribution corresponds to the case where no OSCs are used, i.e.,
here we essentially repeated the analysis by Clemens (1985). The
result obtained is discussed in the next section. Figure 5b
corresponds to the entire set of objects under consideration. The
residual velocity dispersion is 8.8 km s$^{-1}$. The maximum
residual velocity is 45 km s$^{-1}$. As we see from the figure, a
periodic structure of the velocity distribution attributable to
the effect of Galactic spiral density waves is traceable.
Therefore, it is quite natural to use a spectral analysis to
investigate the residual velocities in an effort to obtain
quantitative characteristics.

We can see from our comparison of Figs. 5a and 5b that the fitting
residual velocity curve in Fig. 5b has more distinct peaks; the
similarity of the two periods in the inner Galaxy is better seen
in this curve than in the curve of Fig. 5a.

Figure 6 shows the power spectra of the residual velocities as a
function of the natural logarithm of the wavelength-to-$R_0$ ratio
($\ln \lambda/R_0$). Figure 6a pertains to the residual velocities
$\Delta V_{I}$ obtained by subtracting $V_{rot}-V_0$ (where we
found the velocity $V_0=208$ km s$^{-1}$ above), representing the
deviation from a flat rotation curve, while Fig. 6b pertains to
the residual velocities $\Delta V_{II}$ obtained by subtracting
the derived Galactic rotation curve from the original distribution
of $V_{rot}$ (Fig. 5b).

The power spectra are fairly wide, but we give them only up to
$\ln \lambda/R_0=-3.5$. The amplitude of the spectrum decreases
rapidly with increasing frequency (decreasing wavelength).

We can see from our comparison of Figs. 6a and 6b that taking into
account the derived Galactic rotation curve with parameters (9)
allows us to get rid of the low-frequency spectral component
almost completely.

Three~~ dominant~~ peaks with wavelengths of 2.5, 1.4, and 0.9 kpc
($\ln \lambda/R_0\approx-1.2, -1.8, -2.2$) and amplitudes of 4.7,
2.6, and 3.6 km s$^{-1}$, respectively, can be distinguished in
the power spectrum of the residual velocities shown in Fig. 6b.
The thick line in Fig. 5b indicates the fitting curve of the
residuals constructed using the frequencies of the spectrum in
Fig. 6b for $\ln \lambda/R_0>-2.5$. In contrast to the approach by
Clemens (1985), who estimated the perturbation wavelength and
amplitude only from one dominant peak in the residual power
spectrum, we construct the fitting curve in a wider frequency
range. As a result, we obtain a complex periodic curve that
represents satisfactorily the residuals.

Four main peaks with the following characteristics can be
distinguished in Fig. 5b:

 (1) $R_1=3.9$ kpc, $\Delta V_1= 8.2$ km s$^{-1}$;

 (2) $R_2=5.8$ kpc, $\Delta V_2= 9.2$ km s$^{-1}$;

 (3) $R_3=8.2$ kpc, $\Delta V_3= 1.0$ km s$^{-1}$;

 (4) $R_4=11.4$ kpc, $\Delta V_4=1.8$ km s$^{-1}$.

 The distance differences between the peaks are
 $\Delta R_{21}=1.9$ kpc,
 $\Delta R_{32}=2.4$ kpc and
 $\Delta R_{43}=3.2$ kpc. As can be seen from Fig. 5b, the fourth
peak is revealed with the lowest confidence. In general, Fig. 5b
gives only a qualitative picture of the effect of spiral density
waves. This is because the line of sight passes through different
sections of the spiral arms and, hence, the resulting velocity
distribution is ``smeared''. Nevertheless, we can suggest the
following identification with known spiral arms:
 peak 1 --- the Scutum-Crux arm;
 peak 2 --- the Carina-Sagittarius arm;
 peak 3 --- the Perseus arm;
 peak 4 --- the Norma-Cygnus arm. In general, the
locations of the first three peaks agree well with the two-arm
model by Yuan (1969). Comparison with Figs. 4 and 5 from Russeil
(2003) shows that this identification agrees satisfactorily with
both three-arm and four-arm models of the spiral pattern (see also
Fig. 2 from Vall\'ee (2002)). However, for the best agreement with
models, for example, the model by Russeil (2003), we must assume
that the peak with $R=2.9$~kpc and $\Delta V = 1.9$ km s$^{-1}$ is
significant in our Fig. 5b. In this case, it will be either the
beginning of the Perseus arm for the three-arm model or the
beginning of the Norma-Cygnus arm for the four-arm model.

Figure 7 shows the residual radial velocities of young OSCs as a
function of the Galactocentric distance calculated using Eq. (7).
The power spectrum of the residual radial velocities for young
OSCs obtained using Eq. (8), where the radial velocities are
considered as $\Delta V$, is shown in Fig. 8. The curve of
residuals shown in Fig. 7 has a wavelength $\lambda=1.7$~kpc and
an amplitude of 5.9 km s$^{-1}$; the linear displacement along the
vertical axis is $-1.6$ km s$^{-1}$. We clearly see from our
comparison of Figs. 6b and 8 that, in contrast to the circular
residual velocities, the power spectrum of the residual radial
velocities of young OSCs for wavelengths $\ln \lambda/R_0>-2.5$ is
simpler: in fact, there is only one distinct and symmetric peak
near $\ln\lambda/R_0\approx-1.7$. Therefore, the derived curve of
residuals is nearly sinusoidal in the range $R=6.9$~kpc.

When the circular velocities are analyzed, the error in $\lambda$
is $\pm0.3$ kpc and the error in the perturbation velocity
amplitude is $\pm0.5$ km s$^{-1}$ (504 points). We estimated these
values by the Monte Carlo method. When the radial velocities of
140 young OSCs are analyzed, these errors have the following
values: the error in $\lambda$ is $\pm0.5$~kpc and the error in
the perturbation velocity amplitude is $\pm1.1$ km s$^{-1}$.

\section*{DISCUSSION}
\subsection*{The Galactic Rotation Curve}

Having analyzed the data on hydrogen clouds in a wide $R$ range,
up to $2R_0$, Brand and Blitz (1993) found a nearly flat Galactic
rotation curve ($A=-B$). However, the Oort constants
 $A= 15.50\pm0.27$ km s$^{-1}$ kpc$^{-1}$ and
 $B=-12.17\pm0.67$ km s$^{-1}$ kpc$^{-1}$ we found,
which are in good agreement with the result of various recent
studies (for an overview of the determinations of the Oort
constants, see Bobylev 2004), show that the rotation curve in the
solar neighborhood is nevertheless not flat. The low-frequency
peak in Fig. 6a shows that the amplitude of the deviations of the
Galactic rotation curve from a flat one is about 8 km s$^{-1}$ in
the range 3 kpc $<R<12$~kpc.

Note that, in general, our Galactic rotation curve in the R range
under consideration agrees satisfactorily with the ``composite''
rotation curve by Clemens (1985).

The shape of the rotation curve in the outer Galaxy depends
significantly on the adopted value of $R_0$ (Clemens 1985; Olling
and Merrifield 2000; Zabolotskikh et al. 2002). Our Galactic
rotation curve is intermediate between the two rotation curves
from Olling and Merrifield (2000) constructed using currently
available data for $R_0=7.1$ and 8.5 kpc.

As we see from Fig. 3a, our Galactic rotation curve agrees well,
within the $1\sigma$ confidence interval, with one of the best
(since it was constructed by carefully reconciling the distance
scales of various samples) curves to date constructed by
Zabolotskikh et al. (2002) for $R_0=7.5$~kpc.

As we see from Fig. 3b, our Galactic rotation curve differs
significantly from the rotation curve that was found by Dias and
L\'epine (2005) using data from Clemens (1985) only in the distant
outer part of the Galaxy ($R>11$~kpc). This difference stems from
the fact that the observational data in the outer Galaxy are
scarce, while the available data are unreliable. In this respect,
the unique observational data for S269 (Fig. 3a) are of great
importance.

In general, it can be concluded that our Galactic rotation curve
in the range 3 kpc $<R<12$ kpc under consideration describes well
the observational data used and is in good agreement with the
results of other authors.

\subsection*{The Phase of the Sun in the Spiral Wave}

Having analyzed the velocities of HI clouds in the inner Galaxy,
Clemens (1985) found the deviation of the motion of the local
standard of rest from the circular one with a velocity
 $\Delta V_{rot}=+7.0\pm1.5$ km s$^{-1}$ in the direction of Galactic rotation.

In our view, this motion can also be associated with the effect of
a density wave. Let us consider an example from Rohlfs (1977),
$$
\displaylines{\hfill
      \sigma_1 = |\widehat \sigma| \cos \chi,\hfill\llap(10)\cr\hfill
       V_{rad} =-|f_R| \cos \chi,\hfill\cr\hfill
 \Delta V_{rot}= |f_\theta| \sin\chi~\hfill
}
$$
for the region within the corotation radius in the case of a
tightly wound spiral. Here, $\chi=f(R,\chi_\odot)$ is the phase of
the spiral wave,
 $\chi_\odot$ is the phase of the Sun in the spiral wave,
 $\sigma_1$ is the matter density,
 $\widehat \sigma$ is the density perturbation amplitude,
 $f_R$ and $f_\theta$ are the amplitudes of the radial and tangential velocity perturbation
 components, the center of the spiral arm corresponds to the phase $\chi=0$
 (for details, see the description to Fig. 21 in the
monograph by Rohlfs (1977)).
 For the case of $\Delta V_{rot}=+7.0$ km s$^{-1}$
under consideration, the phase of the solar neighborhood in the
wave will be close to $-\pi/2$. It thus follows that the Sun is
located at the outer edge of the arm.

The solar velocity components $V_\odot(u_\odot,v_\odot,w_\odot)$
we found lead us to conclude that there is a motion of the
centroid of our sample along the Galactic $Y$ axis with a velocity
$\Delta V_{rot}=V_{LSR}-v_\odot=-6.2\pm0.8$ km s$^{-1}$ relative
to the local standard of rest with parameters $(U,V,W)_{LSR} =
(10.00,5.25,7.17)\pm(0.36,0.62,0.38)$ km s$^{-1}$ (Dehnen and
Binney 1998) that differs significantly from zero. Note that
Dehnen and Binney (1998) determined the velocity of the Sun
relative to the local standard of rest based on the proper motions
of $\approx12000$ main-sequence stars from the Hipparcos Catalogue
(1997). Our previous analysis (Bobylev and Bajkova 2007) of the
space velocities for $\approx5000$ F and G dwarfs using radial
velocities showed that
 $(U,V,W)_{LSR} = (8.7,6.2,7.2)\pm(0.5,2.2,0.8)$ km s$^{-1}$.
 Thus, the reliability of the parameters $(U,V,W)_{LSR}$ is beyond doubt.

As we see, the velocity $\Delta V_{rot}$ of the centroid of our
sample along the $Y$ axis is negative. Hence, in this case, the
Sun should be located at the inner edge of the spiral arm, having
a phase close to $\pi/2$. This contradiction between the two
estimates of the phase of the Sun was also pointed out by
Zabolotskikh et al. (2002).

It is clearly seen from Fig. 5a, where we essentially repeated the
analysis of data from Clemens (1985), that the perturbation
velocity in the solar neighborhood is positive, $\Delta V_{rot}=
+2$ km s$^{-1}$. As can be seen from Fig. 5b, including the data
on OSCs in the analysis changes significantly the picture in the
immediate neighborhood of the Sun. The perturbation velocity at
$R=R_0$ is $\Delta V_{rot}=-5.5\pm0.5$ km s$^{-1}$; the phase of
the Sun in the wave should then be close to $\pi/2$.

However, we again run into a significant contradiction. If the
phase of the Sun is assumed to be close to $\pi/2$, then the
radial velocities should decrease with increasing distance from
the Sun in the direction of increasing $R$, but the reverse is
true in Fig. 7.

Note that our OSC residual velocity distributions (Figs. 5b and 7)
are in good agreement with the results of the analyses of the
velocity ($V_R$ and $V_\theta$) fields for OB associations
performed by several authors: see Fig. 3 in Mel'nik et al. (2001)
and Fig. 1 in Mel'nik (2003).

We associate the above contradictions with the influence of the
Orion arm, since a significant fraction of the nearby OSCs (Fig.
1) belong to this structure. In the opinion of several authors
(Yuan 1969; Weaver 1970; Elmegreen 1980; Efremov 1997), the Orion
arm is a branch at the outer edge of the Cygnus spiral arm (Melnik
2003) or a spur. Previously, Bobylev et al. (2007) found that the
centroid of the OSCs belonging to the Orion arm lags behind the
local standard of rest with a velocity of $\approx-10$ km
s$^{-1}$, while the typical lag velocity for the centroids of the
OSC belonging to other arms is $\approx-5$ km s$^{-1}$.

All of the data obtained are difficult to reconcile within the
framework of a simple model. As was shown by Sitnik and Mel'nik
(1999) and Mel'nik (2003, 2006), the real picture of the solar
neighborhood may be much more complex.

Nevertheless, the phase of the Sun in the spiral wave can be
determined from radial velocities. The amplitude of the OSC radial
velocity perturbations is higher than that of the circular
velocity perturbations. The radial velocities are almost free from
the discrepancy in the velocities of the centroid and the local
standard of rest (see the vertical displacement in Fig. 7). The
amplitude of the radial velocity perturbations is essentially
determined by the spiral density wave. Thus, for example, as was
pointed out by Lin et al. (1969), the stars in the solar
neighborhood can go away from their birthplace to distances of
$\sim­1.2$ kpc over a period of $\sim­10$ Myr, which is
attributable mainly to the rotation velocities, while the stars go
away in the radial direction to distances that are a factor of 10
smaller due to the small pitch angle of the spiral arms. Based on
the data in Fig. 7, we conclude that the phase of the Sun in the
density wave is slightly larger than $-\pi/2$ (a more accurate
value is $\chi_\odot=-0.66\pi$). Consequently, the Sun is located
in the interarm space near the outer edge of the
Carina-Sagittarius arm. This location of the Sun agrees well with
the distribution of stars and gas (Vall\'ee 2002; Russeil 2003)
and with predictions of the density-wave theory. Indeed, as we see
from relations (10), the variations in velocity $V_{rad}$ are in
phase with the variations in density $\sigma_1$.

\subsection*{The Scale Factor $\lambda$}

The distance between the neighboring spiral arms along the
Galactic radius vector is usually denoted by $\lambda$;
occasionally, it is called a scale factor or scale length (Mel'nik
et al. 2001).

Clemens (1985) obtained an estimate of $\lambda = 0.22R_0$. For
the value of $R_0 = 8.5$ kpc used by him, this gives $\lambda =
1.9$ kpc, which is in good agreement with our results.

Having analyzed the space velocities of OB associations within 3
kpc of the Sun, Mel'nik et al. (2001) estimated the scale length
of the periodic variations in radial residual velocity components
along the Galactic radius vector to be $\lambda = 2.0\pm0.2$~kpc.

At the same time, we found that the distance between the spiral
arms does not remain constant in the R range under consideration
and is $\Delta R_{i}=1.9,2.4,3.2$~kpc. This is in agreement with
the description of the density wave as a logarithmic spiral, as
distinct from an Archimedean spiral characterized by a constant
distance between the neighboring turns of the spiral. Details of
the discussion on the specific shape of the spiral in the Galaxy
can be found in the monograph by Efremov (1989). The model using
an Archimedean spiral was suggested by Cowie and Rybicki (1982).

Having analyzed the radial velocities of young OSCs, we obtained
an independent estimate of $\lambda=1.7\pm0.5$~kpc from data in
the range of distances 6 kpc$<R<9$ kpc.

\subsection*{The Perturbation Amplitudes}

The perturbation amplitudes of the circular,
 $f_\theta=4.6\pm0.5$ km s$^{-1}$ (the main peak in Fig. 6b), and radial,
 $f_R= 5.9\pm1.1$ km s$^{-1}$ (the main peak in Fig. 8), velocities that we found are in
good agreement with the minimum perturbation amplitude of 5 km
s$^{-1}$ (at the minimum wavelength $\lambda=0.22R_0$) estimated
by Clemens (1985).

Based on formulas from Lin et al. (1069), Burton (1971) estimated
the change in the amplitudes of the perturbation velocities from a
density wave ($f_R$ and $f_\theta$) as a function of $R$ in the
range $0.2R_0 < R < 1.3R_0$. It turned out that for both
velocities, there is a wide peak in the region $0.5R_0<R<0.9R_0$,
where the velocities reach $\approx8$ km s$^{-1}$ ($f_R$ is larger
than $f_\theta$ by $\approx1$ km s$^{-1}$ everywhere), the
velocities are about 4 km s$^{-1}$ near $0.2R_0$ and decrease to
2.5 km s$^{-1}$ near $1.3R_0$. Our results (Fig. 5b) agree well
with the estimates by Burton (1971).

However, it is hard to say that our results are in good agreement
with those of other authors obtained for the solar neighborhood.
Thus, for example, having analyzed OB associations, Mel'nik et al.
(2001) found the velocities of the perturbations from a spiral
density wave to be $f_R = 7\pm1$ km s$^{-1}$ and $f_\theta =
2\pm1$ km s$^{-1}$. Popova and Loktin (2005) obtained completely
different velocities of the perturbations from a spiral density
wave, $f_R = -4\pm5$ km s$^{-1}$ and $f_\theta = 13\pm3$ km
s$^{-1}$, using data on OSCs and OB stars. Based on Cepheids,
Mishurov and Zenina (1999) found $f_R = 3.3\pm1.6$ km s$^{-1}$ and
$f_\theta = -7.9\pm2.0$ km s$^{-1}$. The analysis of data on
Cepheids performed by Popova (2006) showed that the amplitude of
the perturbation velocities, $f_R = -1.8\pm2.5$ km s$^{-1}$ and
$f_\theta = 4.0\pm3.4$ km s$^{-1}$ are small and do not differ
significantly from zero. Zabolotskikh et al. (2002) found
$f_R\approx-7\pm2$ km s$^{-1}$ and $f_\theta\approx-1\pm2$ km
s$^{-1}$ from data on Cepheids and OSCs.

In general, our results lead us to conclude that the curve in Fig.
5b describes well the large-scale (grand design) spiral structure;
there are peculiarities at $R=R_0$ attributable to the kinematic
properties of the Orion arm.

\section*{CONCLUSIONS}

Based on currently available data on the three-dimensional field
of space velocities of young ($\leq50$~Myr) open star clusters and
the radial velocities of HI clouds and star-forming (H II)
regions, we constructed the Galactic rotation curve in the range
of Galactocentric distances 3 kpc $<R<12$ kpc.

At the first step, we determined a smoothed curve that fitted
fairly accurately the Galactic rotation curve in this range of
distances. The curve parameters (9) were found by using the first
six terms of the Taylor expansion of the angular velocity of
Galactic rotation at a given Galactocentric distance of the Sun,
$R_0 = 7.5$ kpc. We found that the centroid of the sample moves
relative to the local standard of rest along the Galactic $Y$ axis
with a velocity of $-6.2\pm0.8$ km s$^{-1}$.

At the second step, we performed a spectral analysis of the
circular velocity residuals of the objects under consideration
from the derived rotation curve, which attributes the observed
periodic residuals to the effect of density waves. Our spectral
analysis showed that the peak with an amplitude of $4.6\pm0.5$ km
s$^{-1}$ corresponding to a wavelength $\lambda = 2.5\pm0.3$ kpc
is dominant. A similar analysis of the radial velocities for young
OSCs confirmed the presence of periodic perturbations from a
density wave with an amplitude of $5.9\pm1.1$ km s$^{-1}$ and a
wavelength $\lambdaл = 1.7\pm0.5$~kpc.

Based on the described approach, Clemens (1985) obtained similar
results, but only for the inner Galaxy  ($R<R_0$), where highly
accurate data on hydrogen are available for tangential points. In
contrast, for the outer Galaxy ($R>R_0$), the radial velocities of
hydrogen (HI and HII) clouds usually have large errors, which
makes it much more difficult to interpret the residuals.

We were able to advance slightly farther than $R_0$, because we
used the total space velocities of OSCs to determine the
projections of their circular velocities. In our view, this
approach improves the homogeneity of the sample.the tangential
point method allows the total space velocity of hydrogen clouds to
be analyzed, since the velocity vector lies entirely on the line
of sight under the assumption of purely circular rotation of
hydrogen at the tangential point. Therefore, supplementing the
sample of objects with OSCs seems natural.

We found that the distance between the spiral arms increases with
Galactocentric distance, being 1.9, 2.4, and 3.2 kpc; this is in
agreement with the description of the density wave as a
logarithmic spiral. The perturbation amplitude also changes --- it
is largest in the inner part of the Galaxy, $\approx9$ km
s$^{-1}$, and decreases to $\approx1$ km s$^{-1}$ in its outer
part.

Analysis of the radial velocities of a sample of young OSCs
suggests that the phase of the Sun in the density wave is close to
$-\pi/2$ and that the Sun is located in the interarm space near
the outer edge of the Carina.Sagittarius arm.

\subsection*{ACKNOWLEDGMENTS}

We wish to thank R.-D. Scholz, who provided the version of the
CRVOCA catalog even before its appearance in the Strasbourg
database, S.V. Lebedeva for help in working with our database, and
the referees for useful remarks that contributed to an improvement
of the paper. This work was supported by the Russian Foundation
for Basic Research (project no. 08-02-00400).

\section*{REFERENCES}
{\small

~~~~ 1. V. S. Avedisova, Astron. Zh. 82, 488 (2005) [Astron. Rep.
49, 435 (2005)].

2. T. M. Bania, Astroph. J. 216, 381 (1977).

3. V. V. Bobylev, Pis'ma Astron. Zh. 30, 185 (2004) [Astron. Lett.
30, 159 (2004)].

4. V. V. Bobylev, A. T. Bajkova, and S. V. Lebedeva, Pis'ma
Astron. Zh. 33, 809 (2007) [Astron. Lett. 33, 720 (2007)].

5. V. V. Bobylev and A. T. Bajkova, Astron. Zh. 84, 418 (2007)
[Astron. Rep. 51, 372 (2007)].

6. J. Brand and L. Blitz, Astron. Astrophys. 275, 67 (1993).

7. W. B. Burton, Astron. Astrophys. 10, 76 (1971).

8. W. B. Burton and M. A. Gordon, Astron. Astrophys. 63, 7 (1978).

9. L. L. Cowie and G. B. Rybicki, Astroph. J. 260, 504 (1982).

10. D. P. Clemens, Astroph. J. 295, 422 (1985).

11. A. K. Dambis, A. M. Mel'nik, and A. S. Rastorguev, Pis'ma
Astron. Zh. 27, 68 (2001) [Astron. Lett. 27, 58 (2001)].

12. W. Dehnen and J. J. Binney, Mon. Not. R. Astron. Soc. 298, 387
(1998).

13. W. S. Dias, B. S. Alessi, A. Moitinho, et al., Astron.
Astrophys. 389, 971 (2002).

14. W. S. Dias and J. R. D. L\'epine, Astroph. J. 629, 825 (2005).

15. Yu.N. Efremov, Sites of Star Formation in Galaxies (Nauka,
Moscow, 1989) [in Russian].

16. Yu. N. Efremov, Pis'ma Astron. Zh. 23, 659 (1997) [Astron.
Lett. 23, 579 (1997)].

17. D.M. Elmegreen, Astroph. J. 242, 528 (1980).

18. M. Fich, L. Blitz, and A. A. Stark, Astroph. J. 342, 272
(1989).

19. The Hipparcos and Tycho Catalogues, ESA SP-1200 (1997).

20. M. Honma, T. Bushimata, Y. K. Choi, et al., astroph/
0709.0820v1 (2007).

21. N. V. Kharchenko, R.-D. Scholz, A. E. Piskunov, et al.,
Astron. Nachr. 328 (2007).

22. P. G. Kulikovskii, Stellar Astronomy (Nauka, Moscow, 1985) [in
Russian].

23. H. S. Liszt and W. B. Burton, Astroph. J. 236, 779 (1980).

24. C. C. Lin and F. H. Shu, Astroph. J. 140, 646 (1964).

25. C. C. Lin, C. Yuan, and F. H. Shu, Astroph. J. 155, 721
(1969).

26. A. M. Mel'nik, A. K. Dambis, and A. S. Rastorguev, Pis'ma
Astron. Zh. 27, 521 (2001) [Astron. Lett. 27, 521 (2001)].

27. A. M. Mel'nik, Pis'ma Astron. Zh. 29, 349 (2003) [Astron.
Lett. 29, 304 (2003)].

28. A. M. Mel'nik, Pis'ma Astron. Zh. 32, 9 (2006) [Astron. Lett.
32, 7 (2006)].

29. Yu. N.Mishurov and I. A. Zenina, Astron. Astrophys. 341, 81
(1999).

30. C. A. Murray, Vectorial Astrometry (Adam Hilger, Bristol,
1983).

31. I. I. Nikiforov, Astron. Zh. 76, 403 (1999) [Astron. Rep. 43,
345 (1999)].

32. K. F. Ogorodnikov, Dynamics of Stellar Systems (Fizmatgiz,
Moscow, 1958; Pergamon, Oxford, 1965).

33. R. P. Olling and M. R. Merrifield, Mon. Not. R. Astron. Soc.
311, 361 (2000).

34. M. E. Popova and A. V. Loktin, Pis'ma Astron. Zh. 31, 743
(2005) [Astron. Lett. 31, 663 (2005)].

35. M. E. Popova, Pis'ma Astron. Zh. 32, 274 (2006) [Astron. Lett.
32, 244 (2006)].

36. A. E. Piskunov, N. V. Kharchenko, S. R\"{o}ser, et al.,
Astron. Astrophys. 445, 545 (2006).

37. A. S. Rastorguev, E. V. Glushkova, A. K. Dambis, and M. V.
Zabolotskikh, Pis'ma Astron. Zh. 25, 689 (1999) [Astron. Lett. 25,
595 (1999)].

38. K. Rohlfs, Lectures on Density Wave Theory (Springer-Verlag,
Berlin, 1977; Mir,Moscow, 1980).

39. D. Russeil, Astron. Astrophys. 397, 133 (2003).

40. M. R. Samal, A. K. Pandey, D. K. Ojha, et al., astroph/
0708.4137v1 (2007).

41. T. G. Sitnik and A. M. Mel'nik, Pis'ma Astron. Zh. 25, 194
(1999) [Astron. Lett. 25, 156 (1999)].

42. Y. Xu,M. J. Reid, X.W. Zheng, et al., Science 311, 54 (2006).

43. X.-X. Xue, H. W. Rix, G. Zhao, et al., astroph/ 0801.1232v1
(2008). 44. J. P. Vall\'ee, Astroph. J. 566, 261 (2002).

45. H. Weaver, in Proceedings of the IAU Symposium: Interstellar
Gas Dynamics, Ed. by H. Habing (Reidel, Dordrecht, 1970), Vol. 39,
p. 22.

46. C. Yuan, Astroph. J. 158, 871 (1969).

47. M. V. Zabolotskikh, A. S. Rastorguev, and A. K. Dambis, Pis'ma
Astron. Zh. 28, 516 (2002) [Astron. Lett. 28, 454 (2002)].

 }

\newpage

\begin{figure}[t]
{
\begin{center}
  \includegraphics[width=80mm]{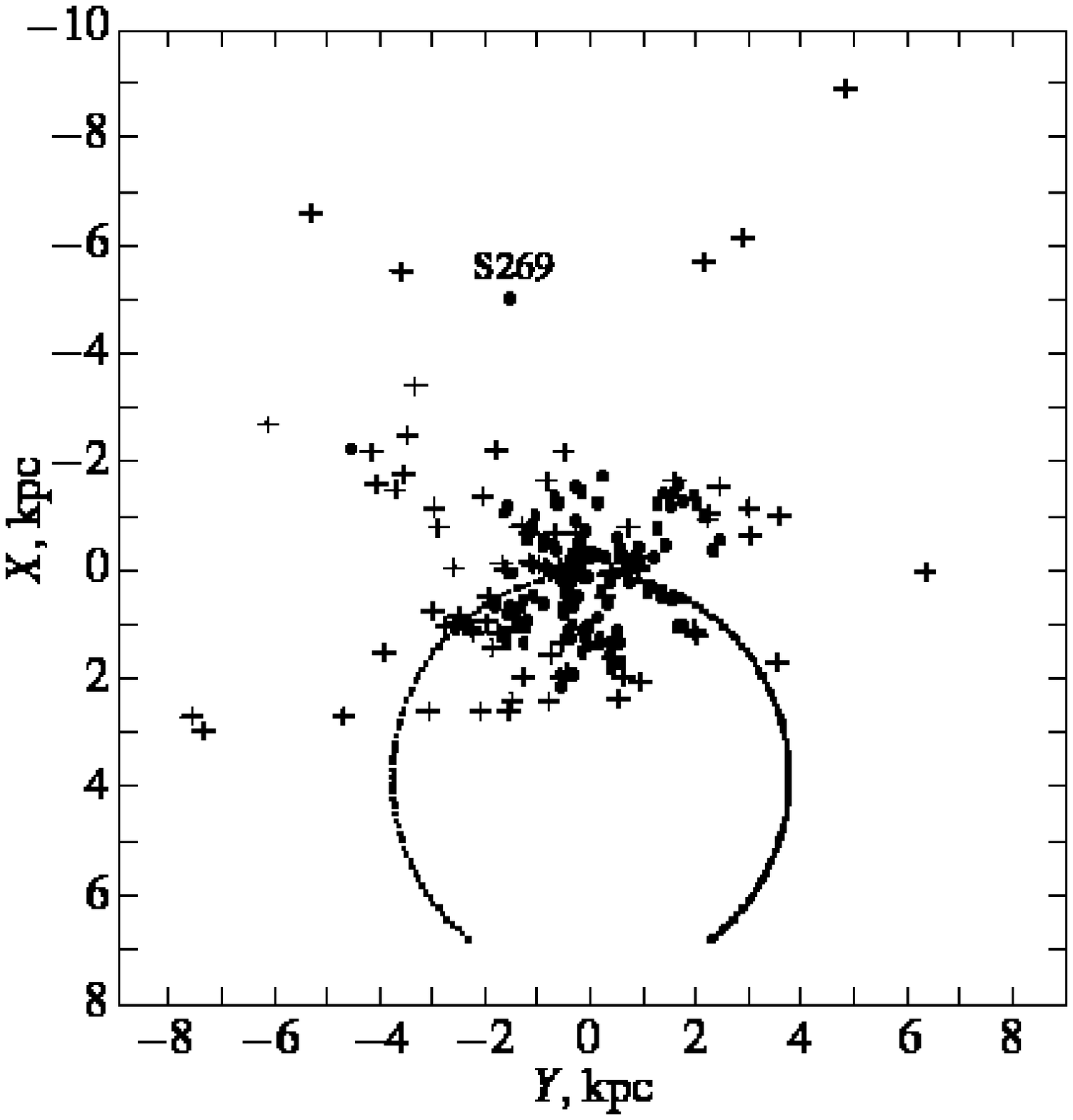}
\end{center}
} Fig.~1. Coordinates of the objects being analyzed in projection
onto the Galactic XY plane. The Sun lies at the coordinate origin.
The pluses, dots, and filled circles denote the HII regions,
tangential points, and OSCs, respectively.
\end{figure}

\begin{figure}[t]
{
\begin{center}
  \includegraphics[width=100mm]{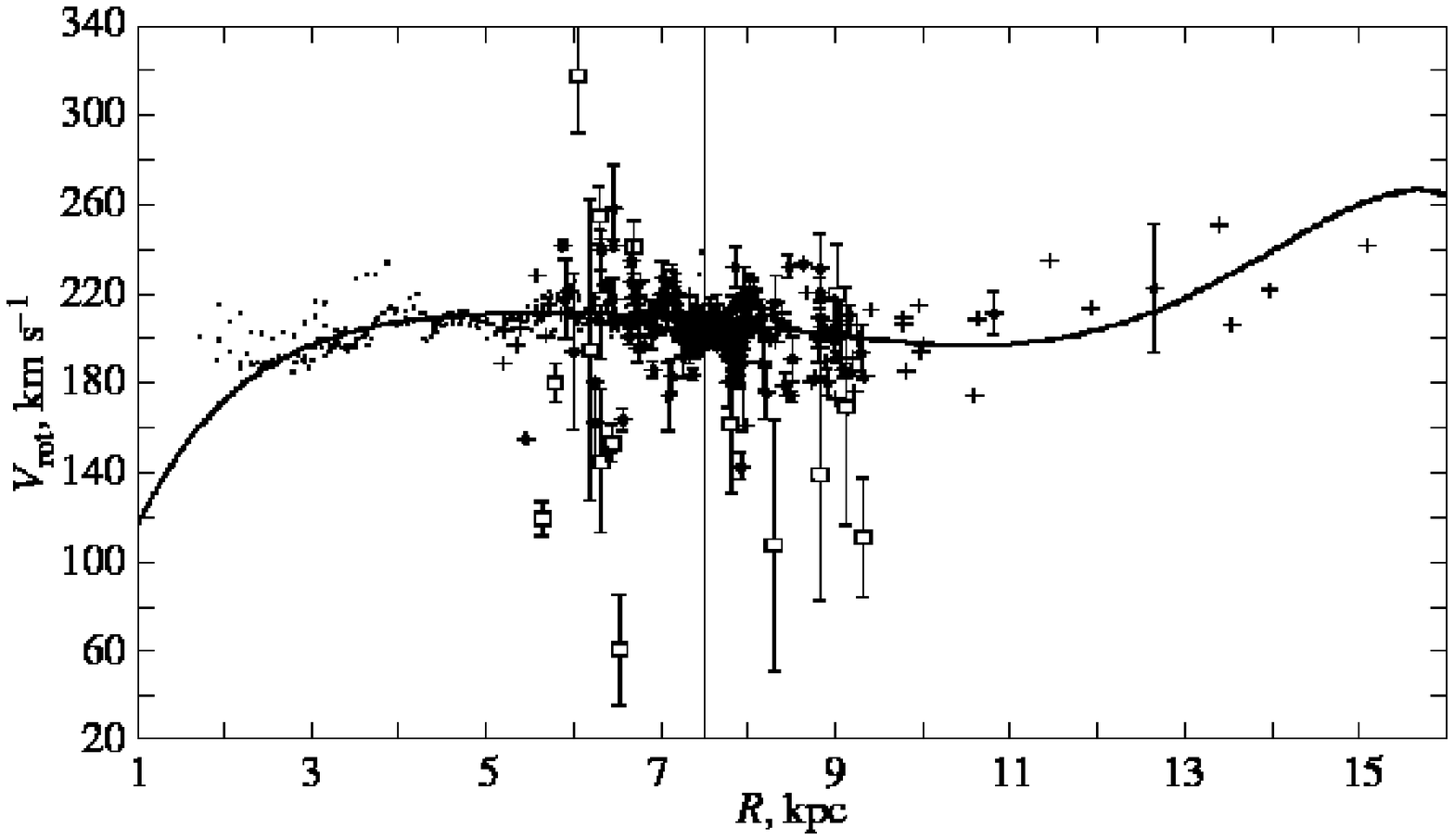}
\end{center}
} Fig.~2. Galactic rotation curve. The vertical line indicates
$R_0=7.5$ kpc. For OSCs, the velocities were calculated only from
the radial velocities using Eq. (5).
\end{figure}

\newpage
\begin{figure}[t]
{
\begin{center}
  \includegraphics[width=140mm]{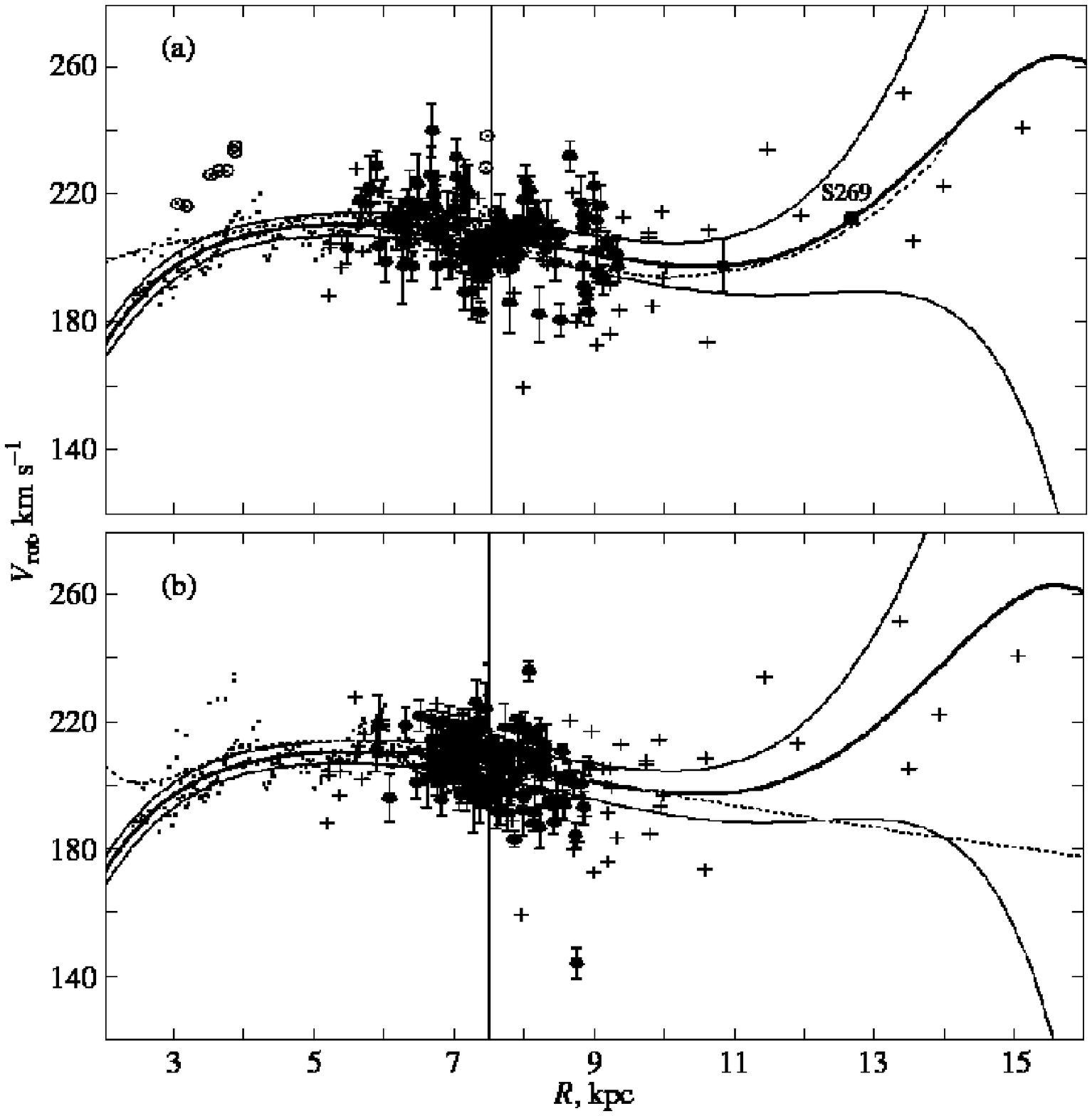}
\end{center}
} Fig.~3. Galactic rotation curve (thick line), the thin lines
mark the boundaries of the $1\sigma$ confidence intervals, the
vertical line indicates $R_0=7.5$~kpc, the symbols $\odot$ in
panel (a) mark the tangential points that were not used to
determine the rotation curve parameters and the remaining symbols
are the same as those in Fig.~1. The dotted line in panels (a) and
(b) denote the rotation curve as derived by Zabolotskikh et al.
(2002) and Dias and L\'epine (2005), respectively.
\end{figure}

\newpage
\begin{figure}[t]
{
\begin{center}
  \includegraphics[width=120mm]{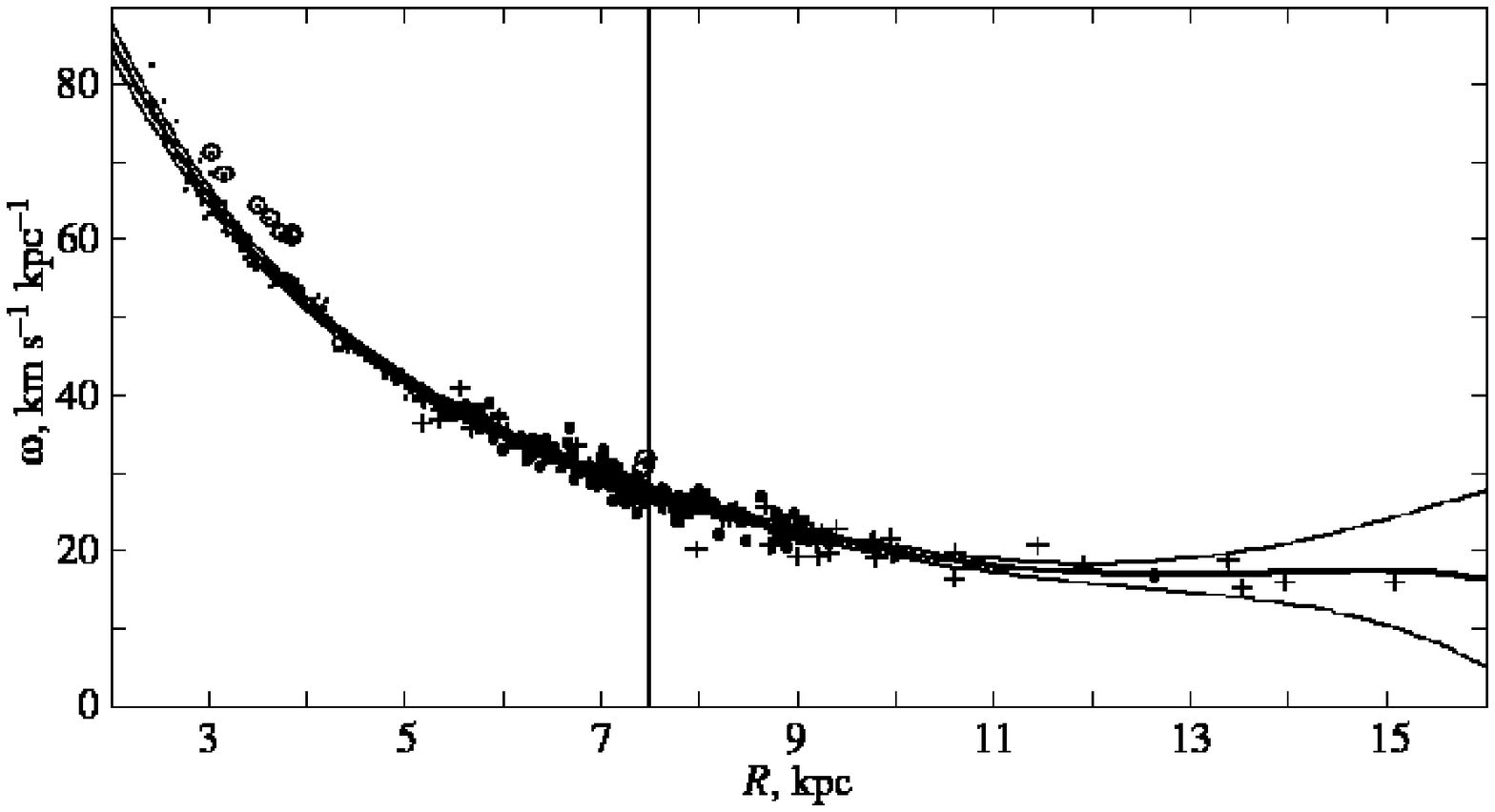}
\end{center}
} Fig.~4. Angular velocity of Galactic rotation versus
Galactocentric distance. The symbols are the same as those in
Figs. 1 and 3, the thin lines mark the boundaries of the $1\sigma$
confidence intervals, and the vertical line indicates
$R_0=7.5$~kpc.
\end{figure}

\newpage
\begin{figure}[t]
{
\begin{center}
  \includegraphics[width=140mm]{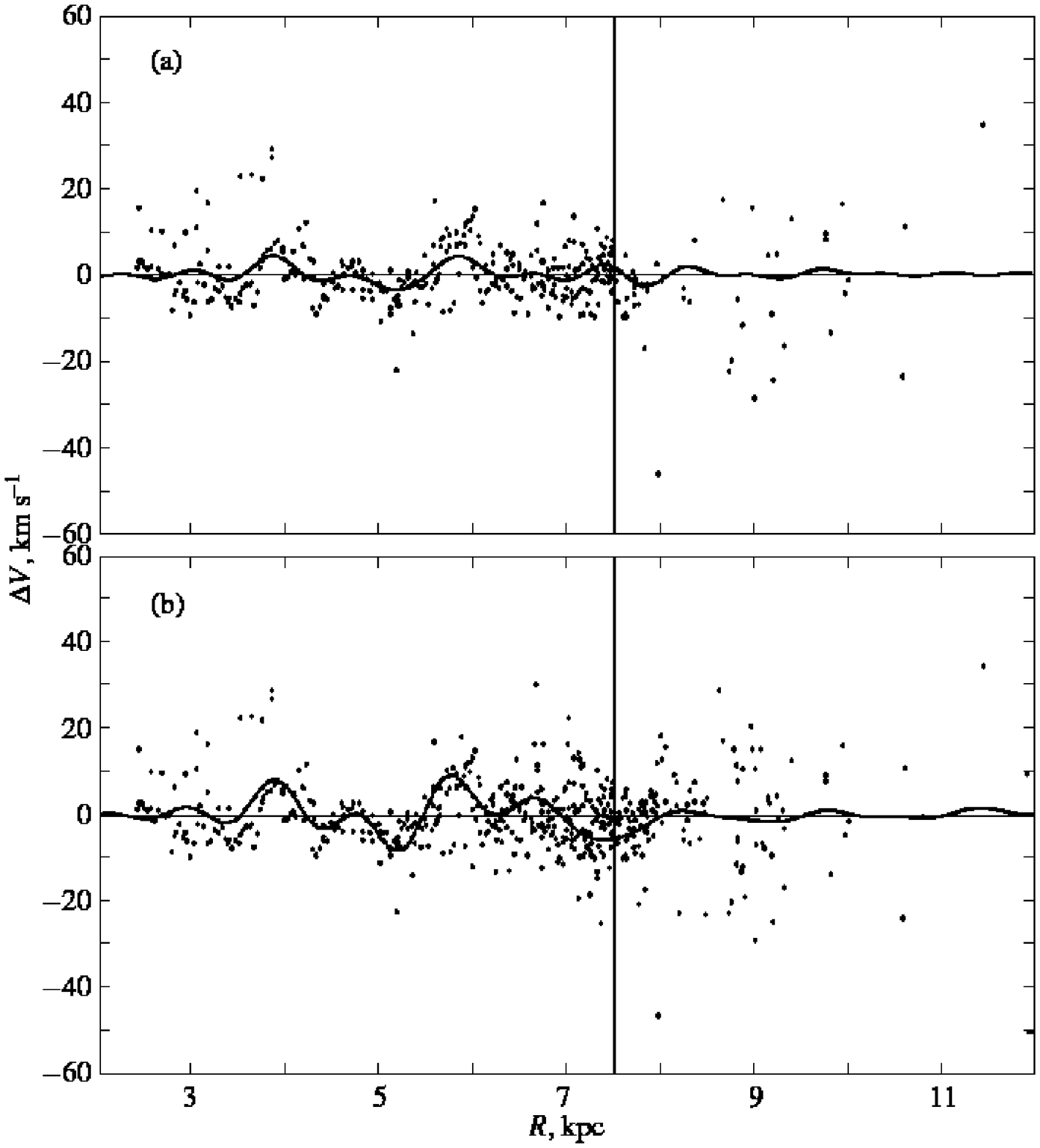}
\end{center}
} Fig.~5. Residual circular velocities versus Galactocentric
distance: (a) no OSCs are used; (b) the entire set of objects
being analyzed is used.
\end{figure}

\newpage
\begin{figure}[t]
{
\begin{center}
  \includegraphics[width=140mm]{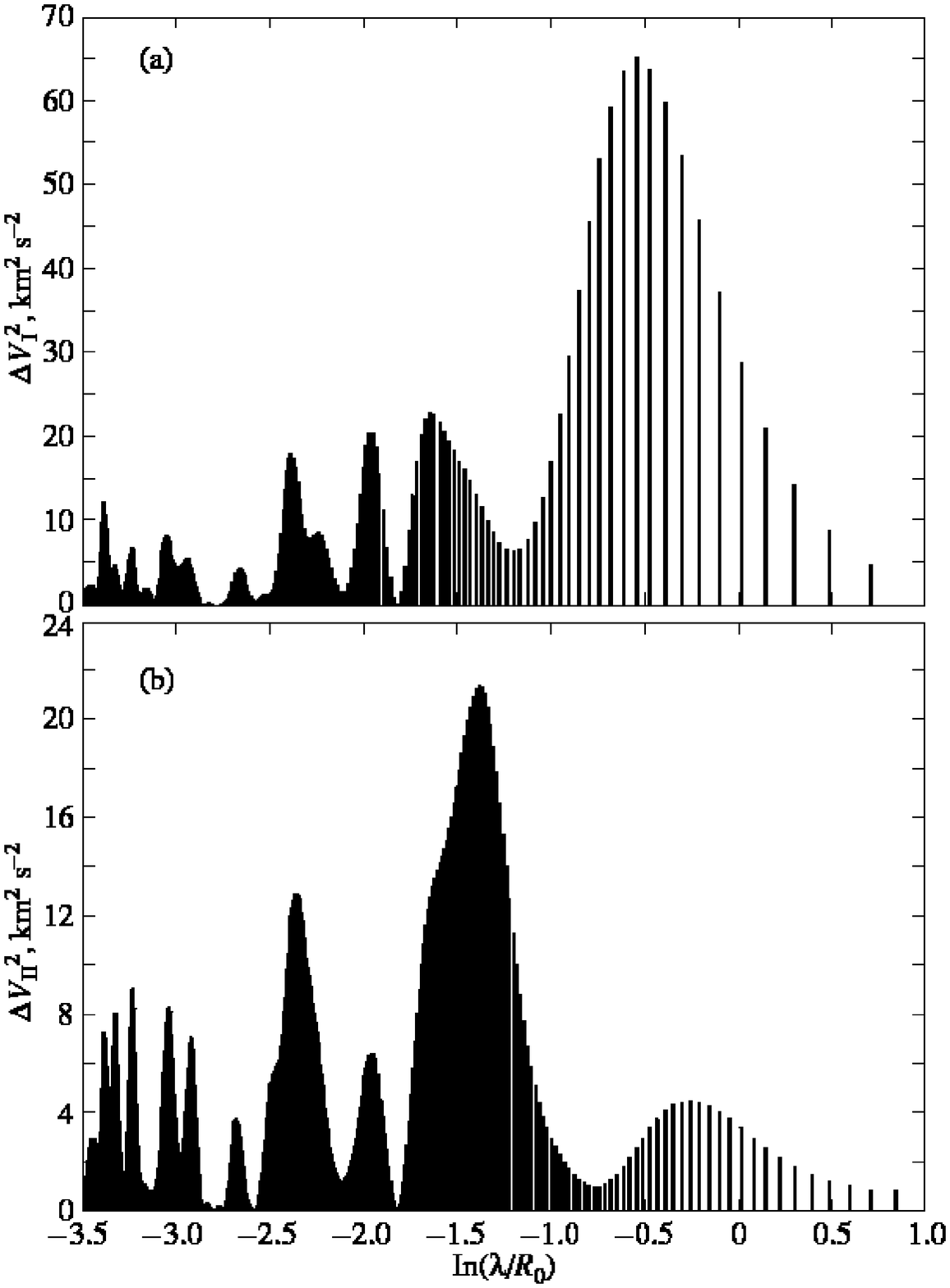}
\end{center}
} Fig.~6. Power spectrum of the residual rotation velocities as a
function of $\ln(\lambda/R_0)$: (a) residual velocities relative
to the flat curve; (b) residual velocities relative to curve (9).
\end{figure}

\newpage
\begin{figure}[t]
{
\begin{center}
  \includegraphics[width=100mm]{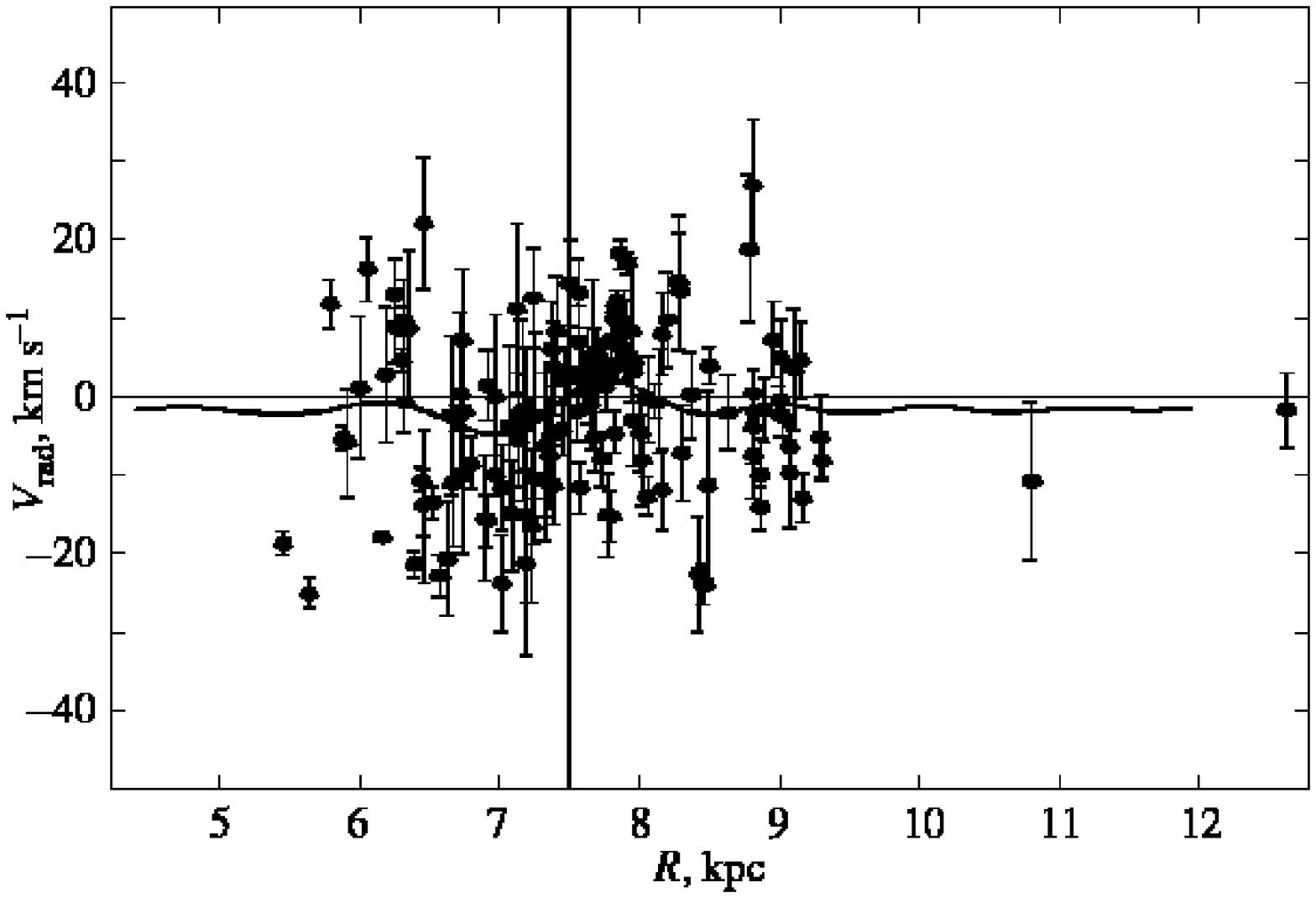}
\end{center}
} Fig.~7. Residual radial velocities of young OSCs versus
Galactocentric distance.
\end{figure}

\newpage
\begin{figure}[t]
{
\begin{center}
  \includegraphics[width=100mm]{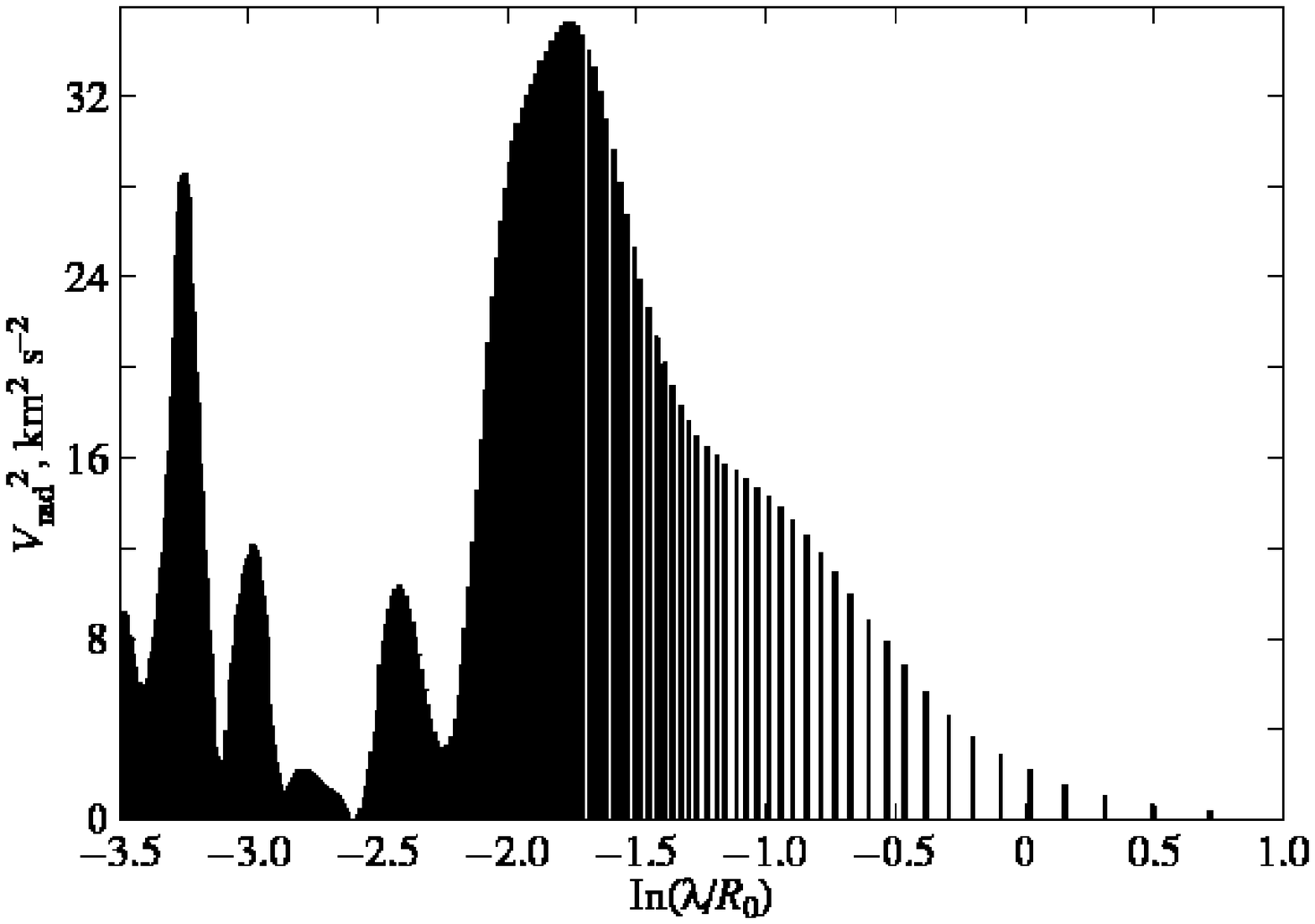}
\end{center}
} Fig.~8. Power spectrum of the residual radial velocities for
young OSCs as a function of $\ln(\lambda/R_0)$.
\end{figure}

\end{document}